\documentclass{ws-mplb}
\usepackage{amssymb,caption}
\usepackage{cmbright}       
\usepackage{enumerate}
\usepackage{rotating} 

\usepackage{longtable}

\usepackage{amsfonts}

\newcounter{eqletter}

\def\mathletters{%

\setcounter{eqletter}{0}%

\addtocounter{equation}{1}

\edef\curreqno{\arabic{equation}}

\edef\@currentlabel{\theequation}

\def\theequation{%

\addtocounter{eqletter}{1}\thechapter.\curreqno\alph{eqletter}%

}%

}

\begin{document}

\markboth{Kai P. Schmidt and G\"otz S. Uhrig}
{Spectral Properties of Magnetic
Excitations in Cuprate Two-Leg Ladder Systems}

%
\catchline{}{}{}{}{}
%

\title{Spectral Properties of Magnetic Excitations\\
in Cuprate Two-Leg Ladder Systems}

\author{\footnotesize Kai P. Schmidt}

\address{Institute of Theoretical Physics, 
\'{E}cole Polytechnique F\'{e}d\'{e}rale de Lausanne\\
CH 1015 Lausanne, Switzerland\\
kaiphillip.schmidt@epfl.ch}

\author{G\"otz S. Uhrig}

\address{Theoretische Physik, Geb.\ 38, FR 7.1, 
Universit\"at des Saarlandes\\
D-66123 Saarbr\"ucken, Germany\\
uhrig@lusi.uni-sb.de}

\maketitle

\begin{history}
\received{(received date)}

\revised{(revised date)}



\end{history}

\begin{abstract}
This article summarizes and extends
the recent developments in the microscopic modeling of the
magnetic excitations in cuprate two-leg ladder systems. The microscopic
Hamiltonian comprises dominant Heisenberg exchange terms plus an additional
four-spin interaction which is about five times smaller. We give an overview 
over the relevant energies like the one-triplon dispersion, 
the energies of two-triplon bound states and the positions of multi-triplon 
continua and  over relevant spectral properties like spectral
weights and spectral densities in the parameter regime appropriate for
cuprate systems. It is concluded that an almost complete understanding of the
magnetic excitations in undoped cuprate ladders has been obtained as measured 
by inelastic neutron scattering, inelastic light (Raman) scattering
 and infrared absorption.
\end{abstract}

\keywords{magnetic excitations, spin ladders, high-$T_{\rm c}$-superconductivity, spin liquids, four-spin ring exchange, spectral densities, continuous unitary transformations}

\section{Introduction}

The origin of  high-temperature superconductivity in layered cuprates
is still heavily debated. Nevertheless, it has become evident
in the last years that the important physics takes place in the two-dimensional
copper-oxide planes\cite{ander87,ander97}. 
At zero doping, these compounds are  insulating long-range ordered
antiferromagnets which are described by
(extended) Heisenberg models\cite{manou91}. On doping,
they can be viewed as doped Mott insulators\cite{orens00}.
Doping destroys the antiferromagnetic order rapidly.
At some finite value of doping, the superconducting state is realized.

The long-range ordered N\'eel state at zero doping suggests to base
a microscopic description on magnons which interact with the doped holes.
But the holes destroy the ordered N\'eel state giving rise to
a spin-liquid state with short range antiferromagnetic correlations.
Thus, it would be more appropriate to set up a microscopic description in
terms of the elementary excitations of the spin-liquid state, which can
be fractional $S=1/2$ spinons\cite{fadde81} or integer $S=1$ 
triplons\cite{schmi03c}. Their interplay with the doped holes is expected
to be crucial for the understanding of the pairing mechanism.
But a quantitative theory is so far not in sight.

There is a different class of systems which are spin liquids already
in the undoped case due to their lower dimensionality\cite{barne93,dagot96},
namely the cuprate two-leg ladders. Two-leg ladders have a finite 
spin gap$^{9-12}$
and the elementary
excitations are triplons with $S=1$. The ground states of these ladders
can be viewed as realizations of the so-called resonance valence bond (RVB) 
state, a coherent superposition of singlet dimers,
proposed by Anderson for the two-dimensional  cuprates\cite{ander87,liang88}.
Thus, the physical understanding of the cuprate ladders as doped spin
liquids is an important step to the understanding of the
two-dimensional cuprates.

There has been decisive progress in the understanding of
undoped cuprate ladders both in theory$^{7,8,11,12,14,15-26}$
and in experiment$^{27-37}$.
It is found that a microscopic modeling of the
magnetic degrees of freedom has to include a four-spin interaction besides the
usual nearest-neighbor Heisenberg exchange. 
The size of the four-spin exchange in cuprate
ladders is determined to be about 20\% of the leading nearest-neighbour 
Heisenberg exchange$^{17,24,30-32}$. 

The present article has two objectives. First, it reviews briefly
 our current understanding of the undoped cuprate ladders. 
Second, it extends the results for  two-triplon bound states and spectral
densities to the experimentally relevant model including the
 four-spin ring exchange.

The article is set up as follows.  In Sect.\ \ref{sect:model}, we introduce the
cuprate ladder compounds as well as the microscopic model and we  give a short 
overview of the recent developments. Then we present  a sketch of the
method we use to describe the two-leg ladder system. In the sequel, the
major properties of the two-leg ladder system are explored in the 
parameter regime relevant for the cuprate ladders. In Sect.\
\ref{sec:sw_ladder_4sp} the spectral weights are discussed. Sect.\
\ref{sect:energy} describes the energy properties, namely the one-triplon
dispersion, the energies of the two-triplon bound states and the
positions of the multi-triplon continua. The last two parts,
 Sects.\ \ref{sd:S1} and \ref{sec_lad_SD_S0}, deal with important spectral
 densities. The focus is laid on the dynamic structure factor which is
 relevant for inelastic neutron scattering (INS). A short summary of inelastic
 light scattering, i.e., Raman spectroscopy, and infrared absorption (IR absorption) is also provided. Finally, the article is summarized and concluded in Sect.\ 
\ref{sect:sum}

\section{Cuprate Ladders and the Microscopic Model}
\label{sect:model}
\subsection{Materials}
\setcounter{equation}{0}

Realizations of two-leg ladders are established in SrCu$_2$O$_3$
(Sr123) \cite{hiroi91,uehar96} and A$_{14}$Cu$_{24}$O$_{41}$ (A14) with
A=\{Sr,Ca,La\} \cite{mccar88,siegr88} compounds. The fundamental building
block in both compounds are edge-sharing copper-oxide plaquettes where
adjacent copper ions are linked linearly to one another by 
intercalated oxygen ions. In A14 corner-sharing plaquettes occur also;
they give rise to spin chains. But the corresponding energy scale is much 
lower and we do not focus on them here.
Sr$_{0.4}$Ca$_{13.6}$Cu$_{24}$O$_{41}$ becomes superconducting 
for high pressure\cite{uehar96} enhancing the interest in 
this class of compounds in particular.

\begin{figure*}
  \begin{center}
     \includegraphics[width=\textwidth]{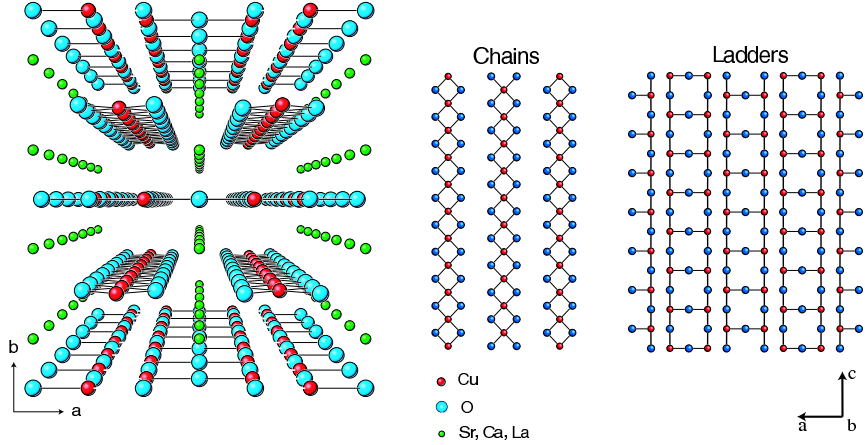} 
\end{center}
    \caption{(color online) Schematic view of the telephone number compound 
A14. The left  panel is a three-dimensional view in the layered material.
There are planes of ladder structures (plane at the bottom, in the middle and
at the top), of chain structure (the third and the seventh plane) and the
planes containing the A cations. The top views of the
ladder plane (Cu$_2$O$_3$) and of the chain plane (CuO$_2$) are shown in the 
panels on the right side.} 
\label{fig:ladder_a14cu24o41}
\end{figure*}

SrCu$_2$O$_3$ is the prototype of a system of
weakly coupled Cu$_2$O$_3$ spin-ladders. The
relevant orbitals of the copper atoms are the planar $d_{x^2-y^2}$ orbitals. 
The superexchange runs via the $p_x$ or $p_y$ orbitals of the
 intermediate oxygen\cite{ander50}. The interladder coupling is
significantly smaller because the superexchange via a Cu-O-Cu path with a 
90$^\circ$ angle is reduced due to the vanishing overlap of the 
orbitals$^{42-44}$.
In addition, the interladder coupling is frustrated because 
the spin on one ladder is coupled equally to two adjacent spins on the
neighboring ladder, which themselves are coupled antiferromagnetically,
see the topview of a plane of ladders on the right side in Fig.\ 
\ref{fig:ladder_a14cu24o41}. The disadvantage of SrCu$_2$O$_3$ is
that it has to be grown under very high pressure so that only small single
crystals or polycrystals are available\cite{loffe02}. 
This hampers the momentum resolution and the counting statistics
of scattering experiments, in particular of neutron scattering investigations.

The second class of compounds are the so-called telephone-number compounds
A$_{14}$Cu$_{24}$O$_{41}$ with A=\{Sr,Ca,La\}, see left panel
in   Fig.\ \ref{fig:ladder_a14cu24o41} \cite{ammer00b}.
Two sorts of planes occur. One sort of planes consists of 
two-leg Cu$_2$O$_3$ ladders as described in the last paragraph.
The other sort consists of corner-sharing CuO$_2$
chains. Both one-dimensional structures, ladders and chains, respectively,
are oriented along the c-axis. The two sorts of planes are
illustrated on the right hand side of Fig.\ 
\ref{fig:ladder_a14cu24o41}. It is instructive to write
A$_{14}$Cu$_{24}$O$_{41}$ $=$ (A$_{2}$Cu$_{2}$O$_{3}$)$_7$(CuO$_2$)$_{10}$ in
order to emphasize the existence of two structures. 
This formula assumes a commensurate ratio of seven rungs in the ladders
matching ten copper sites in the chains. But a closer investigation
reveals that this is only an approximation;
the ratio is in fact incommensurate, see e.g.\  Ref.\ 47
and references therein.

Ladders and chains form two-dimensional
layers which are stacked in b-direction in alternation with 
A$ \in \{$Sr, Ca, La $\}$ layers,
see left panel in Fig.\ \ref{fig:ladder_a14cu24o41}. 
Large single  crystals of A14 can be grown fairly easily so that  
inelastic neutron scattering investigations of high resolution are possible.

The distance between two copper atoms is roughly the same in rung and in leg 
direction. The typical order of magnitude of the nearest-neighbor Heisenberg 
coupling is 1000 K. It is similar to the one in the two-dimensional 
cuprates forming square lattices\cite{eccle98,colde01b}.
The coupling  between two copper atoms on neighboring ladders is 
weakly ferromagnetic due to the bonding angle of 90$^\circ$. $^{42,44}$
 The nearest-neighbor exchange coupling in the 
chains is mediated via two symmetric Cu-O-Cu bonds with about 90$^\circ$ 
bonding angle implying a weakly ferromagnetic coupling. 
Its typical value is an order of magnitude smaller than the couplings
 in the ladders\cite{eccle98}. 
The difference in energy scales makes it possible to study
the behavior of both substructures, ladders and chains, separately.

Two telephone number compounds, namely Sr$_{14}$Cu$_{24}$O$_{41}$
(Sr14) and La$_{6}$Ca$_8$Cu$_{24}$O$_{41}$ (La6Ca8) are the focus of this work.
 The system Sr14 is intrinsically doped with 6 holes per unit cell. Using X-ray
 absorption spectroscopy, it is possible to conclude that there are on average 
0.8 holes in the ladders and 5.2 holes in the chains\cite{nucke00} which can be
 explained by a higher electronegativity in the chains\cite{mizun97}.

It is possible to substitute Sr$^{2+}$ by isovalent Ca$^{2+}$ or by 
trivalent La$^{3+}$ or Y$^{3+}$. 
The substitution by La$^{3+}$ reduces the number of holes in 
the system. The limiting case of an undoped sample is reached
by the (formal) compound La6Ca8. So far, crystal growth has not been
able to realize  La6Ca8. But 
there exist crystals up to La$_{5.2}$Ca$_{8.8}$Cu$_{24}$O$_{41}$ 
which can already be viewed as (almost) undoped samples\cite{windt01,gruni02b}.

\subsection{Minimal Microscopic Model}

We want to focus here on spin isotropic models which
do not contain anisotropies. This focus is justified by the fact that
the anisotropies in cuprate systems are generically very small relative
to the dominant spin isotropic exchange couplings along linear 
Cu-O-Cu bonds.
Most often,  Heisenberg exchange couplings are considered in the description 
of the magnetic properties of insulating electronic systems. 
It has been realized early by Dirac\cite{dirac29} and it was later introduced 
in condensed matter physics by Thouless\cite{thoul65} that the general 
expression for the exchange Hamiltonian reads
\begin{equation}
 H_{\rm ex}=\sum_{n=2}^\infty \sum_{\alpha_n} 
J_{\alpha_n} (-1)^{{ P}_{\alpha_n}}\Pi_{\alpha_n}\quad ,
\end{equation} 
where $\Pi_{\alpha_n}$ denotes a permutation operator of the
localized spins. The sum runs over 
all permutations labelled by $\alpha_n$ of $n$ spins. 
${P}_{\alpha_n}$ is the 
parity of the permutation $\Pi_{\alpha_n}$. From the physics point of view,
these permutations
correspond to the exchange of $n$ spins. The first term represents the 
well-known  two-particle exchange which is usually the dominant contribution.

The importance of the exchange processes with more than two particles was 
realized first in the description of the magnetism of solid $^3$He 
\cite{roger83}. In $^3$He,  the two-particle exchange is small due to a 
hardcore repulsion of the atoms leading to steric blocking.
Thus, higher ring exchange processes are important.

The situation in cuprate systems is different to solid $^3$He because
the dominant exchange is the nearest-neighbor Heisenberg exchange. 
Moreover, the typical geometries consist of square plaquettes so that the
lowest order ring exchange comprises four spins.
To be precise, we denote the
Hamiltonian of the two-leg ladder with additional four-spin ring exchange 
by
\begin{eqnarray}
\label{eq:HamiltonianP}
 H^{\rm p} &=& \frac{J_\perp^{\rm p}}{2} \sum\limits_i \Pi_{1,i;2,i}
+ \frac{J_\parallel^{\rm p}}{2} \sum_{\tau,i} \Pi_{\tau,i;\tau,i+1}
+H_{\rm cyc}^{\rm p} \\
H_{\rm cyc}^{\rm p} &=& \frac{J_{\rm cyc}^{\rm p}}{4}
\sum_{<\tilde i\tilde j\tilde k\tilde l>}
\left( \Pi_{\tilde i\tilde j\tilde k\tilde l}+
\Pi_{\tilde i\tilde j\tilde k\tilde l}^{-1}\right) \ .\nonumber
\end{eqnarray}
Here $\Pi_{\tilde i;\tilde j}$ is the permutation operator of two-particles,
$\tau\in\{1,2\}$ labels the legs and $i$ the rungs. The composite index 
$\tilde i= (i,\tau)$ combines the leg and the rung information. Then the 
permutation operator of four spins is $\Pi_{\tilde i\tilde j\tilde k\tilde l}$ 
where  $<{\tilde i\tilde j\tilde k\tilde l}>$  restricts the sum to four spins 
on a plaquette. We use the subscript  $_{\rm cyc}$ from `cyclic' to indicate 
quantities which refer to ring exchange processes.

It is convenient to introduce spin operators and to rewrite 
(\ref{eq:HamiltonianP}) in terms of  a Heisenberg model plus additional 
four-spin interactions\cite{brehm99}
\begin{eqnarray}
\label{eq:Hamiltonian}
 H &=& J_\perp \sum\limits_i {\bf S}_{1,i}
{\bf S}_{2,i} + J_\parallel \sum_{\tau,i} {\bf S}_{\tau,i} {\bf S}_{\tau,i+1}
+H_{\rm cyc}\\
\nonumber
H_{\rm cyc} &=& J_{\rm cyc}\sum_{\rm plaquettes}\big[({\bf S}_{1,i}{\bf
  S}_{1,i+1})({\bf S}_{2,i}{\bf S}_{2,i+1}) + ({\bf S}_{1,i}{\bf
  S}_{2,i})({\bf S}_{1,i+1}{\bf S}_{2,i+1})\\
\nonumber
 && \qquad 
- ({\bf S}_{1,i}{\bf S}_{2,i+1})({\bf S}_{1,i+1}{\bf S}_{2,i})\big]\quad .
\end{eqnarray}
Both Hamiltonians (Eq.\ \ref{eq:HamiltonianP} and Eq.\ \ref{eq:Hamiltonian}) 
are the same up to two-spin interactions along the diagonal of the spin ladder.
 It is known that these couplings are small 
(of the order of $0.03 J_\perp$ \cite{mizun99})
 so that we restrict the discussion in the following to the 
Hamiltonian in Eq.\ \ref{eq:Hamiltonian}. Note that both Hamiltonians
 are in use in the literature. 
The relation between the exchange parameters of both 
Hamiltonians reads
\begin{eqnarray}
 J_\perp &=& J_\perp^{\rm p}+\frac{1}{2}J_{\rm cyc}^{\rm p} \\
 J_\parallel &=& J_\parallel^{\rm p}+\frac{1}{4} J_{\rm cyc}^{\rm p}\\
 J_{\rm cyc} &=& J_{\rm cyc}^{\rm p}\quad .
\end{eqnarray} 

In agreement with the above general considerations, evidence for significant 
ring exchange processes in high temperature superconductors has been 
found\cite{roger89,schmi90b} very soon after their discovery\cite{bedno86}. 
It was observed that the derivation of effective
spin models from the underlying Hubbard models implies that the most
important \emph{corrections} to the nearest-neighbor Heisenberg exchange 
are four-spin ring exchange terms. This has been substantiated in 
a sequence of papers considering either the three-band Hubbard 
model\cite{mizun99,mulle02a,calza03}
or the single-band Hubbard model$^{59-61}$. 
The effects of the ring exchange on the Raman response have 
been discussed very early\cite{roger89,schmi90b,honda93}.

For planar cuprates, a first experimental indication for the relevance of the 
four-spin ring exchange was conjectured on the basis of the line shape of IR absorption data\cite{loren99c}. An unambiguous  experimental signature was found by INS in La$_2$CuO$_4$ in 2001\cite{colde01b}. The  spin-wave dispersion can be explained only by the inclusion of four-spin interactions. The Raman line shape also points towards a significant four-spin exchange 
coupling\cite{katan02a,katan03}. Recently, it was shown that the magnetic 
excitations in the stripe-ordered phase of La$_{1.875}$Ba$_{0.125}$CuO$_4$ are 
quantitatively described assuming  20\% ring exchange coupling\cite{uhrig04a}.

Indications for four-spin interactions in cuprate ladders was found in 1999
by INS\cite{brehm99,matsu00a,matsu00b} for La6Ca8. The fits of
the experimentally measured one-triplon dispersion by the results for the
standard Heisenberg model yielded $J_\parallel/J_\perp\approx 2$
where $J_\parallel$ is the coupling along the legs and 
$J_\perp$ the one along the rungs of the ladder. The very large
deviation of this ratio from unity appears astounding
in view of the geometrical structure of the cuprate ladders
which is fairly isotropic. The inclusion of the four-spin
ring exchange can resolve this discrepancy. But the 
INS data is not precise enough at higher energies
to determine the  value for the four-spin interaction quantitatively. 

The interaction with light provides a convenient way to obtain information
on energetically higher lying excitations\cite{jurec00,windt01}. 
In particular, bound states can be detected. The information on these
excitations in turn makes a quantitative determination of the ring exchange
coupling possible\cite{nunne02,haga03}.  For La6Ca8 it was found that
$J_{\rm cyc} \approx (0.2-0.25) J_\perp$, $J_\parallel/J_\perp\approx 1.2-1.3$,
and  $J_\perp = 1000-1200$cm$^{-1}$ yields a consistent description of
the available exerimental data\cite{nunne02}. 
Thus the Hamiltonian (\ref{eq:HamiltonianP})
or the Hamiltonian (\ref{eq:Hamiltonian}) constitute the minimal model for
the physics of undoped cuprate ladders.

The analysis of the shape of Raman lines is another means to determine
the parameters of the minimal model\cite{freit00b,schmi01}. It reveals that
the exchange couplings in Sr14 \cite{gozar01} and Sr123 \cite{gossl03} differ 
slightly from those in La6Ca8.
Both Sr14 and Sr123 are described by $x:=J_\parallel/J_\perp\approx 1.4-1.5$ 
and  $x_{\rm cyc}:=J_{\rm cyc}/J_\perp\approx 0.2-0.25$ \cite{schmi05a}. 
The values for Sr14 are consistent 
 with those for IR absorption data for this compound\cite{windt03,gruni05}.

In the preceding paragraphs, we have focussed on the progress in the
experimental analysis. In parallel, there has been important
progress in the theoretical understanding of  spin ladders.
The single-triplon properties were elucidated 
early\cite{barne93,dagot96,eder98}.
The low-lying spectrum in the limit of two coupled chains 
($J_\perp \ll J_\parallel $) has been analyzed by bosonization
\cite{shelt96}. Without ring exchange the spin gap $\Delta$ vanishes linearly 
with $J_\perp$ which is confirmed also by quantum Monte Carlo\cite{greve96}.
Furthermore, the bosonic field theory predicts a singlet state between the 
triplon state and the continuum.

The opposite limit ($J_\perp \gg J_\parallel $) is the local limit of the
system which is also gapped\cite{barne93}. Combined with
the field theoretic and the QMC results we know that  
no phase transition takes place for $0 < x < \infty$.
Moreover, inspection of the interaction of two triplons
in the limit $x\to 0$ reveals that they attract each other
so that bound states occur$^{14-16,73}$.  
The most tightly bound state is always found in the $S=0$ 
sector\cite{uhrig96b,sushk98,jurec00,kotov99}. Important progress in 
high order series expansion for multi-particle states$^{19,21,74-78}$
made it possible to follow these bound states
to the isotropic regime $J_\perp\approx J_\parallel$. 
Besides the energies of the bound states also two-triplon
spectral densities can now be computed by this 
approach\cite{knett01b,schmi01,zheng03a}.

For theoretical completeness, the phase diagram was investigated for
all possible ratios of the coupling constants. If the four-spin interaction
is strongly increased the gap in the spin ladder vanishes at a critical
value\cite{nerse97}. Beyond this value,  the character of the phase changes. 
The ground-state
is no longer the so-called rung-singlet  phase\cite{lauch03,grits03}
which is dominated by singlets on the rungs and which is smoothly connected
to the phase of isolated rungs. The phase for stronger
four-spin interaction is characterized by a staggered dimerization on the legs 
of the ladder$^{80,82-86}$.
Apart from these two phases, a variety of other phases can be reached for
other ratios of $J_{\rm cyc}$ and $J_\perp = J_\parallel$ including
negative values\cite{lauch03,grits03,momoi03,hikih03}. 
We do not discuss the phases other than the rung-singlet phase in more detail 
because the generic experimental systems known so far provide spin ladders in 
the range of the rung-singlet phase only. In view of the ample theoretical 
results for phases beyond the rung-singlet phase it would be exciting if 
experimental realizations for such phases could be found.

\section{Method}
\label{sect:method}

In this part, the essential points concerning the \emph{c}ontinuous
\emph{u}nitary \emph{t}ransformations (CUTs) are discussed$^{75,89-91}$.
Here we use a particle-conserving perturbative 
CUT\cite{knett03a,knett04b,schmi04b} which uses the states on isolated rungs as
reference. The Hamiltonian of a spin ladder including additional four-spin 
interaction (Eq.\ \ref{eq:Hamiltonian}) is re-expressed by
\begin{equation}
\label{H_pert_Lad_4sp}
  \frac{H(x)}{J_{\perp}}=H_{\perp}+xH_{\parallel}+x_{\rm cyc}H_{\rm cyc}
\end{equation}
with $x=J_{\parallel}/J_{\perp}$ and $x_{\rm cyc}=J_{\rm cyc}/J_{\perp}$ as
perturbation parameters.

The CUT is defined by 
\begin{equation}
 \partial_l H (l) = [\eta (l),H(l)]
\end{equation} 
to transform $H(l=0)=H$ from its initial
form (\ref{H_pert_Lad_4sp}) to an effective Hamiltonian
$H_{\rm eff}:=H(l=\infty)$ which \emph{conserves} the number
of elementary triplet excitations on the rungs, the so-called
triplons\cite{schmi03c}. Formally, the conservation of
triplons is expressed by $0= [H_\perp,H_{\rm eff}]$. An appropriate
choice of the infinitesimal generator $\eta$ is given by the matrix
elements  
\begin{equation}
\label{eq:generator}
 \eta_{i,j} (l)={\rm sgn}\left(q_i-q_j\right)H_{i,j}(l)
\end{equation}
in an eigen basis of $H_\perp$; the $q_i$ are the corresponding eigen
values\cite{knett00a}.

The perturbative realiziation of the CUT provides a series expansion for
the various parts of the effective Hamiltonian:
the one-triplon hopping amplitudes and the 
triplon-triplon interaction. 
We have calculated the hopping amplitudes up to order 11 and the
triplon-triplon interaction up to order $10$ in $x$ and $x_{\rm cyc}$.
Since the expansion is done in the couplings which link the rungs 
each additional order increments the range of the processes which
can be captured. So the order characterizes approximately up to which
range processes are included in the calculation. The range is measured
in units of the lattice constant along the legs.

Next, the observables of interest in the
ladder system have to be evaluated  in order to determine the spectral 
properties of the two-leg ladder.
The observables $\mathcal{O}$ are transformed by the
{\emph same} CUT 
\begin{equation}
 \partial_l \mathcal{O} (l) = [\eta (l),\mathcal{O}(l)]\quad .
\end{equation}
The four {\it local} operators considered are
  \begin{eqnarray}
    \label{o1_Lad}
    {\mathcal O}^{\rm I}(r) &=& {\mathbf S}_{1,r} {\mathbf
    S}_{2,r}={\mathcal T}^{\rm I}_0 \\
    \label{o2_Lad}
    {\mathcal O}^{\rm II}_l(r) &=& {\mathbf S}_{l,r}
      {\mathbf S}_{l,r+1}\\\nonumber
      &=&\frac{1}{4}\left( {\mathcal T}_{-2}+{\mathcal
      T}_0+{\mathcal T}_2+{\mathcal T}^{\rm II}_{-1}+{\mathcal
      T}^{\rm II}_1 \right)\\  
    \label{o3_Lad}
    {\mathcal O}^{\rm III}(r) &=& {\mathbf S}^{z}_{1,r}-{\mathbf
      S}^{z}_{2,r} = {\mathcal T}^{\rm III}_{-1}+ {\mathcal T}^{\rm
      III}_1\\ 
    \label{o4_Lad}
    {\mathcal O}^{\rm IV}(r) &=& {\mathbf S}^{z}_{1,r}+{\mathbf
      S}^{z}_{2,r} = {\mathcal T}^{\rm IV}_0 \ .
  \end{eqnarray}
Here ${\mathcal T}_n$ are local operators changing the triplon number by
$n$. The explicit action of these operators is given
elsewhere\cite{knett04b}. The index $l=1,2$ in Eq.\ \ref{o2_Lad} denotes the
leg on which the observable operates.
In contrast to the triplon-conserving effective Hamiltonian $H_{\rm eff}$, the
effective observables also excite or annihilate triplons. 
The parts requiring that an excitation is present before
the application of the observable do not matter at zero temperature.
 We have calculated those parts
of the effective observables which correspond to the creation
and annihilation of one and two triplons up to order $10$.
The part creating  three triplons is determined 
up to order $8$ and the four-triplon parts up to
 order $7$ in $x$ and $x_{\rm cyc}$.

As it is common in physics the symmetry properties of the
injected triplons play an important role. Here we discuss
spin and parity. The first two observables ${\mathcal
  O}^{\rm I}$ and ${\mathcal O}^{\rm II}$ excite triplons with total spin
zero. Thus they are  relevant for optical experiments like
Raman scattering and IR absorption. The latter two observables inject triplons with total 
spin one. They are important to study the dynamic 
structure factor relevant for INS experiments.

\begin{figure}[t]
  \begin{center}
    \includegraphics[width=8cm]{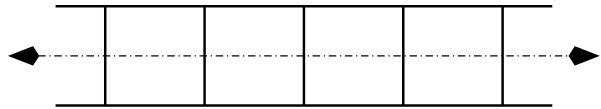}
    \caption{\label{Centerline_Lad}The operator ${\mathcal P}$ reflects
      about the depicted axis. A {\it single} singlet  on a rung is odd;
      a triplet on a rung is even with respect to ${\mathcal P}$. 
      The action of
      ${\mathcal P}$ on the rung-singlet ground state is defined to be
      of even parity ${\mathcal P}|0\rangle=|0\rangle$, i.e.\ we
      assume an even number of rungs.
      If one singlet is substituted by a triplon in the triplon vacuum  
      $|0\rangle$, the resulting state $|1\rangle$ is odd:
      ${\mathcal P}|1\rangle=-|1\rangle$.} 
  \end{center}
\end{figure}

The parity which is conserved in this ladder system is 
defined for the reflection ${\mathcal P}$ 
about  the center-line of the ladder, see Fig.\ \ref{Centerline_Lad}.
If $|n\rangle$ denotes a state where $n$ rung singlets are
excited to triplons while all other rungs remain in their singlet
state one finds ${\mathcal P}|n\rangle = (-1)^n|n\rangle$, see also caption
of Fig.\ \ref{Centerline_Lad}. The state $|n\rangle$ is  a linear
combination of many $n$-triplon states. So no generality is lost in
writing 
\begin{equation}
  \label{O_d_n}
  {\mathcal O}_{\rm eff}|0\rangle = \sum_{n\ge 0}|n\rangle\ .
\end{equation}
The parity with respect to ${\mathcal P}$ of the observables 
introduced in Eqs.\ 11-14 is clear from their definition:
${\mathcal O}^{\rm III}$ is odd while ${\mathcal O}^{\rm I}$ and
${\mathcal O}^{\rm IV}$ are even. The
 symmetrized observable ${\mathcal O}^{\rm II}=({\mathcal
  O}^{\rm II}_{1}+ {\mathcal O}^{\rm II}_{2})/2$ is equally even. 
The parity
is conserved in the CUT so that ${\mathcal P}$ applied to both sides
of Eq.\ (\ref{O_d_n}) requires
\begin{equation}
  \label{sym_Lad}
  {\mathcal O}_{\rm eff}|0\rangle=\left\{
  \begin{array}{cc}
    \sum_{n}|2n\rangle, & {\mathcal O}_{\rm
    eff}\mbox{ even}\\ 
    \sum_{n}|2n+1\rangle, & {\mathcal O}_{\rm
    eff}\mbox{ odd}
  \end{array}  \right.\ .
\end{equation}
An even (odd) parity of ${\mathcal O}_{\rm eff}$ implies 
that ${\mathcal O}_{\rm eff}$ injects an even (odd) number of triplons into 
the system. The
importance of these symmetries will be discussed in more detail
in Sects.\ \ref{sd:S1} and \ref{sec_lad_SD_S0} where
the experimentally relevant results are presented.

The series for the effective Hamiltonian and for the effective observables 
have to be extrapolated in order to obtain reliable results  describing 
the regime of cuprate ladders. The extrapolation is performed
using the one-triplon spin gap  $\Delta$ as internal 
parameter\cite{schmi03d,schmi04b,schmi03a}.
Usually a standard $[n,2]$ dlog-Pad\'{e} extrapolant or  a $[n,2]$ Pad\'{e} 
extrapolant is used for the  series expressed in $\Delta$.
The achieved range of reliability depends on the quantity under study.
The spin gap can be computed down to its zero. The matrix elements of
the interaction are more difficult to extrapolate. For 
very low values of $x_{\rm cyc}$ the ratio $x$ can be taken up to 1.5. 
For $x_{\rm cyc}$ around $0.2$ the value of $x$ should not exceed 1.2 to 1.4.

\section{Spectral Weights}
\label{sec:sw_ladder_4sp}

The  analysis of  the spectral weights helps to determine the
quality of the description in terms of a small number of triplons. 
The chosen description works well, if
the weight resides mainly in a small number of channels with a
small number of triplons .
If, however, the weight is distributed over a large number of
channels the description is less well adapted to the problem.
 The focus is laid here on  the
spectral weights for realistic values of the four-spin interaction. 

\subsection{$S=1$}

\begin{figure}[t]
\centerline{\psfig{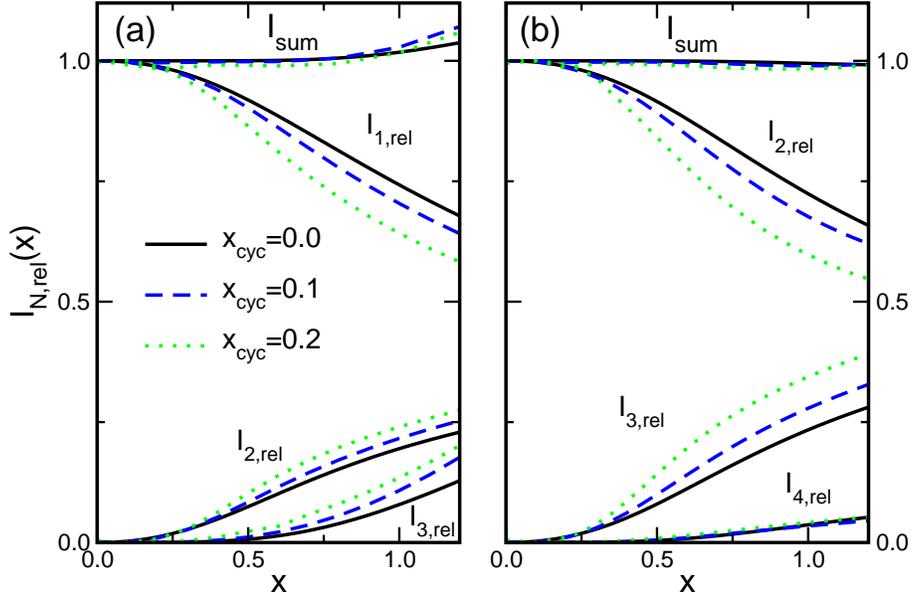}}
\caption{(color online) (a) Relative weights $I_{N,\rm rel}:=
  I_{N}/I_{\rm tot}$ for the $S=1$ operator
  $\frac{1}{2}({\mathcal O}^{\rm III}(r)+ {\mathcal O}^{\rm IV}(r))$ (Eq.\
  \ref{o3_Lad} and Eq.\ \ref{o4_Lad}). The $I_{N}$ are calculated up
  to and including order 10, 10, and 9 in $x$ for $N=$1, 2, and 3,
  respectively. The total intensity $I_{\rm tot}=\sum_{N=1}^\infty I_N$ 
  is rigorosly equal to 1/4. $I_{\rm  sum}=(I_1+I_2+I_3)/I_{\rm tot}=
  I_{1,\rm rel}+I_{2,\rm rel}+I_{3,\rm rel}$ denotes the 
  sum of all plotted contributions; it is to be compared to unity.
  (b) Relative weights for the $S=0$ operator
  ${\bf S}_{1,i}{\bf S}_{1,i+1}$ (Eq.\ \ref{o2_Lad}). The $I_{N}$ are calculated
  up to and including order 10, 8,
  and 7 in $x$ for $N=$2, 3, and 4, respectively. The total
  intensity $I_{\rm tot}$ has been extracted from the 11$^{\rm
    th}$ order result for the ground state energy per spin.  $I_{\rm
    sum}=(I_2+I_3+I_4)/I_{\rm tot} = 
  I_{2,\rm rel}+I_{3,\rm rel}+I_{4,\rm rel}$ denotes the sum of all plotted
  contributions.\label{fig:sw_4sp_S0}
\label{fig:sw_4sp}}
\end{figure}

First, the spectral weights for excitations with total spin $S=1$ are 
discussed. 
The local observable considered is the combination of Eq.\ 
\ref{o3_Lad} and Eq.\ \ref{o4_Lad}
\begin{equation}
 {\mathcal O}^{\rm S=1}(r)=\frac{1}{2}({\mathcal O}^{\rm III}(r)
+ {\mathcal O}^{\rm IV}(r)) \quad .
\end{equation}    
This combination corresponds to the spin operator acting on the spin on leg 1.
The spectral weights were calculated up to order 10 in the one- and two-triplon
channel, up to order 9 in the three-triplon channel. The total spectral weight
$I_{\rm tot}$ is equal to 1/4 for $S=1$ because $(S^z)^2=1/4$.

In Fig.\ \ref{fig:sw_4sp}a the relative spectral weights $I_{N,\rm rel}$ are 
shown for $x_{\rm cyc}=0$ (black solid lines), $x_{\rm cyc}=0.1$ ((blue) 
dashed  lines) and $x_{\rm cyc}=0.2$ (gray (green) dotted lines) and 
$N\in\{1,2,3\}$. In 
addition, the sum of all plotted relative weights $I_{\rm sum}=(I_1+I_2+I_3)/
I_{\rm tot}$ is depicted. The sum is very close to unity. Hence
contributions with more than three triplons only have 
negligible weight in the parameter regime displayed.

The three-triplon sector is extrapolated by a $[5,0]$ dlog-Pad\'{e} extrapolant 
because other extrapolants are spoiled by spurious poles. 
We presume that the depicted extrapolant of the three-triplon weight  
is affected by the vicinity of such poles so that it overshoots
for larger  values of $x$. This conclusion is supported by the observation 
that  the sum of all plotted relative weights overshoots in the same fashion 
as the three-triplon weight increases for increasing $x$. 
The extrapolations of the one- 
and two-triplon sector are very stable so that it is unlikely that they
induce the violation of the sum rule for large $x$.

The general tendency on increasing the four-spin interactions
is the reduction of the one-triplon weight and the increase of the weights for 
two and three triplons. 
Extrapolating the one-triplon spectral weight by dlogPad\'e extrapolants,
one recognizes poles which indicate that the one-triplon weight vanishes at 
the  phase  transition\cite{schmi03d}.
The analysis of $I_2$ in the vicinity of the phase transition indicates
that it stays finite at an approximate value of 25\% . So the larger part
of the weight must have been transferred to channels with three and more
triplons. We conclude that the description in terms of rung triplons
becomes inappropriate on approaching the phase transition to the 
staggered dimerized phase$^{80,83-86}$.
For comparison, we mention that the description in terms of triplons
on dimers for dimerized and frustrated spin chains appears to be much more
robust\cite{schmi03c,schmi04a}.

Fig.\ \ref{fig:sw_4sp}a shows that for realistic values of the four-spin 
interaction the one- and the two-triplon sector contain by far most of the 
spectral weight. Therefore, the dynamic
structure factor for these two channels provides 
an almost complete description of INS experiments on cuprate ladders. 
The corresponding calculation is presented in Sect.\ \ref{sd:S1}. 

\subsection{S=0}

For the $S=0$ case, the local observable ${\mathcal O}_{\rm eff}^{\rm II}$ in 
Eq.\ \ref{o2_Lad} is  considered. The total spectral weight $I_{\rm tot}(x)$ is
calculated reliably
 from the ground state energy per spin $\epsilon_0(x)$ \cite{knett04b}.

In Fig.\ \ref{fig:sw_4sp}b the relative spectral weights $I_{N,\rm rel}$ are
shown for $x_{\rm cyc}=0$ (black solid lines), $x_{\rm cyc}=0.1$ ((blue) dashed
 lines) and $x_{\rm cyc}=0.2$ (gray (green) dotted lines) and $N\in\{
2,3,4\}$. In addition, the sum $I_{\rm sum}=(I_2+I_3+I_4)/I_{\rm tot}$ of all
plotted relative weights is depicted. The sum rule is fulfilled
quantitatively for all values of $x_{\rm cyc}$ depicted in Fig.\ 
\ref{fig:sw_4sp}b.

Since a triplon has $S=1$ there can be no one-triplon contribution $I_1$ to the
$S=0$ response. The $I_2$ contribution is the leading one. Similar to the $S=1$
case, the leading contribution loses spectral weight upon switching on 
$x_{\rm cyc}$. Simultaneously, triplon sectors with more than two triplons gain
spectral weight. The two-triplon sector contains only  about 50\% of the 
spectral weight for realistic parameters 
($x_{\rm cyc}\approx 0.2, x\gtrapprox 1$). This means that there are sizable
contributions from channels with more than two triplons. This is seen in 
optical experiments\cite{gruni03a} where three-triplon contributions are 
important. In addition, the four-triplon sector matters also, giving rise to
finite life-times (see discussion below for Fig.\ 
\ref{fig:ladder_continua_MT}).

In summary, the analysis of the spectral weights shows that the description 
where rung triplons are taken as quasi-particles works very well. In the regime
of parameters which matter for cuprate ladders, the sectors with low numbers of
triplons capture the essential physics. Sectors with more triplons play a 
certain, but minor, role.

\section{Energy Properties}
\label{sect:energy}

\begin{figure}
  \begin{center}
    \includegraphics[width=12cm]{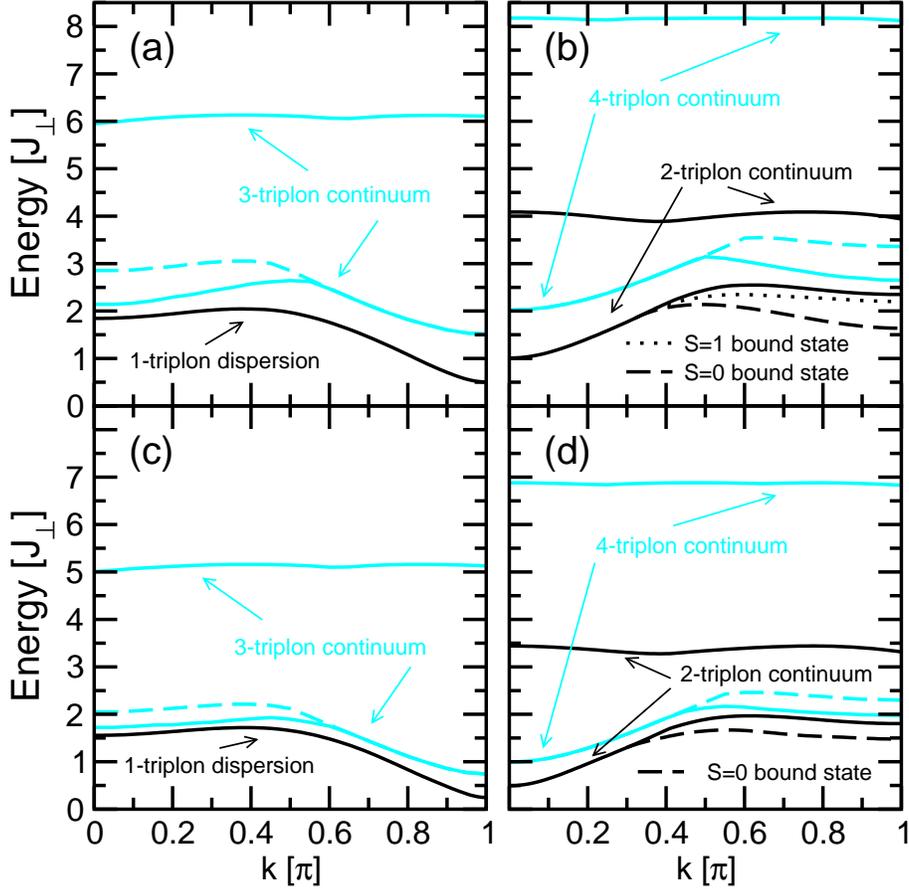}
    \caption{(color online) Dispersions and multi-triplon energies of the 
      two-leg ladder for 
      $x=1$. The ring exchange is zero in panels (a) and (b) while it is  
      $x_{\rm cyc}=0.2$  in panels (c) and (d). One- and three-triplon energies
      are shown in the left panels while two- and four-triplon energies are 
      depicted in the right panels. Recall that the sectors of even and of odd 
      number of triplons do not couple.
      The gray (cyan) solid lines correspond to the lower and to the 
      upper band edge of the multi-triplon continua including two-triplon 
      binding effects. The dashed gray (cyan) lines 
      denote the lower band edges of these continua neglecting 
      two-triplon binding.}
    \label{fig:ladder_continua_MT}
\end{center}
\end{figure}

The purpose of this section is to illustrate the energetic properties of the
two-leg ladder with four-spin interaction in the rung-singlet phase. The 
influence of the four-spin interactions on the
relevant energies is investigated. The focus is laid on the one-triplon
dispersion, on the energies of the two-triplon bound states and on the position
of the multi-triplon continua. In Fig.\ \ref{fig:ladder_continua_MT}
the most relevant energies are shown for $x=1$ with $x_{\rm cyc}=0$ and
$x_{\rm cyc}=0.2$. The results found are qualitatively 
generic for $x=1.0-1.5$ and  $x_{\rm cyc}=0-0.25$.

\subsection{One-Triplon Dispersion}
\label{ssec:disp_4sp}

The one-triplon dispersion can be calculated by very many techniques ranging 
from exact diagonalization over density matrix renormalization 
and quantum Monte Carlo to diagrammatic techniques
and series expansions, see e.g.\ Ref.\ 7.
For consistency we show and discuss the series expansion results here. 
In the left panel of Fig.\ \ref{fig:ladder_continua_MT},
the one-triplon dispersion is shown for $x=1$ with $x_{\rm cyc}=0$ 
(upper panels) and $x_{\rm  cyc}=0.2$ (lower panels). 
The one-triplon dispersion diplays a global minimum at $k=\pi$
which defines the one-triplon gap $\Delta$. 
Additionally, a local minimum exists at $k=0$. 
This minimum can be understood as a crossover property to the
physics of spin-chains where the dispersion at $k=0$ and at $k=\pi$
are degenerate for the infinite system.
It is caused by level repulsion between the one-triplon state and
the three-triplon continuum which approaches the one-triplon state
more and more as $x\to \infty$.  The three-triplon continuum is displayed as 
gray (cyan) lines in the left panels of Fig.\ \ref{fig:ladder_continua_MT}. 

The main effect of the four-spin interaction is a global shift of the
one-triplon dispersion to lower energies. The overall shape of the dispersion
is almost unchanged. The gap is reduced considerably by about a factor of 2 for
$x_{\rm cyc}=0.2$. This property in particular explains the importance of the 
four-spin interaction for a quantitative description of cuprate ladders.

\subsection{Two-Triplon Continuum and Bound States}
\label{ssec:2tc_4sp}
Regarding two-triplon energies not only the one-particle kinetics of the two 
triplons but also the two-triplon interaction is 
important\cite{trebs00,knett01b,knett04b}. Generally, one has a two-triplon 
continuum and two-triplon bound states for a given total momentum. 
The two-triplon energies are shown as black lines in the right panels
of Fig.\ \ref{fig:ladder_continua_MT}.

There are two regimes. For small momenta $k<0.35\pi$, bound states are absent
and a wide two-triplon continuum is present. For larger momenta $k>0.35\pi$ and
$x_{\rm cyc}=0$, there are two bound states below the two-triplon
continuum. There is one bound state with total spin $S=1$ and one with total
spin $S=0$. The binding energy of the $S=0$ bound state is by a factor 2 to 4 
larger than the binding energy of the $S=1$ bound state which is expected
in an antiferromagnet. The dispersion of both bound
states has a maximum at $k\approx 0.5\pi$ and a minimum at $k=\pi$.

Since a finite four-spin exchange shifts the one-triplon dispersion to lower 
energies, also the two-triplon energies are shifted globally to lower 
energies. The shape of the lower and the upper edge of the two-triplon 
continuum remains almost unchanged by the four-spin exchange
because the band edges are solely determined by the one-triplon dispersion.

The  effect of the four-spin interaction on the binding energy of the 
two-triplon bound states (black dashed lines in Fig.\ 
\ref{fig:ladder_continua_MT}) 
is most interesting. The four-spin interaction reduces the
attractive interaction between the triplons. Thus the binding energy of the
bound states is reduced upon increasing $x_{\rm cyc}$ \cite{nunne02}. This
effect is present for total spin $S=0$ and for total spin $S=1$. 
The effect is more pronounced in the $S=1$ case because the $S=1$ two-triplon 
bound state is less tightly bound. As a consequence, the $S=1$ 
bound state is absent for $x_{\rm cyc}=0.2$. The disappearence of the $S=1$
 bound state has interesting effects on the spectral line shape of
the two-triplon contribution to the dynamic structure factor which will be
discussed in Sect.\ \ref{sd:S1}. 

\subsection{Multi-Triplon Continua}
Next the relative positions of the multi-triplon continua are
investigated. The determination of the multi-triplon band edges in
combination with the multi-triplon spectral weights (see Sect.\
\ref{sec:sw_ladder_4sp}) is important to estimate possible life-time effects
which are neglected in the extrapolation procedure. 
Recall that the generator (\ref{eq:generator}) orders the states according
to their number of triplons. This works perfectly if the energy of a state
with $n+2$ triplons has a higher energy than all states with $n$ or less
triplons\cite{knett00a,mielk98}. 
But this is not always case, see e.g.\ Fig.\ \ref{fig:ladder_continua_MT}.
If the correlation between triplon number and energy is not valid 
the state with less triplons can decay into the state with more
triplons. This hybridization implies an additional broadening of the
spectral densities. It can be viewed as a finite life-time effect.
In this regime, the perturbative CUT based on the generator 
(\ref{eq:generator}) works only approximately. The life-time effects
are neglected.

For simplicity, the $n$-triplon interactions with $n>2$ (Ref.\ 77) 
are not included in the calculation of the band edges. They are 
expected to be very small\cite{kirsc04}.

The reduction of the one-triplon gap $\Delta$ by the four-spin interaction 
brings all multi-triplon continua closer together in energy. 
The lower band edge of the three-triplon continuum is very close to 
the one-triplon dispersion for $k\in[0,0.4\pi]$ (left panels). The value for 
the edge of the three-triplon continuum at $k=0$ is the sum of the one-triplon 
gap and the $S=0$ two-triplon bound state energy at $k=\pi$.

The situation for two and four triplons is depicted in the right panels of
Fig.\ \ref{fig:ladder_continua_MT}. Both
continua overlap strongly for both parameter sets. The overlap of the 
two- and four-triplon continuum is independent of the total spin. This is so because the minimum four-triplon energy is always a sum of either two times a scattering state of two triplons or the sum of a two-triplon bound state with $S=0$ and a scattering state of two triplons. But the total spectral weight in the four-triplon channel with $S=1$ is smaller than the one with $S=0$ and the resulting life-time effects are expected to be smaller for $S=1$. Possible effects of the neglected processes are discussed later in the sections dealing with spectral densities and with the interpretation of optical experiments.
     
\section{Dynamic Structure Factor}
\label{sd:S1}

This part deals with the dynamic structure factor
 of the two-leg spin ladder with
four-spin interaction. Results for the one- and the two-triplon
contribution which capture most of the spectral weight (see Sect.\
\ref{sec:sw_ladder_4sp}) are shown. The obtained spectral densities are 
directly relevant for INS experiments.

In the discussion of the parity as defined in Sect.\ \ref{sect:method}, 
we found that the observable
$\mathcal{O}^{\rm III}$ has an odd parity while the observable
$\mathcal{O}^{\rm IV}$ has an even parity. Thus, the observable
$\mathcal{O}^{\rm III}$ excites only an odd number of triplons and the
 observable $\mathcal{O}^{\rm IV}$ excites an even number of triplons. 
It follows that the by far most important one- and two-triplon contributions 
can be measured \emph{independently} by INS because the two contributions have 
a different parity. In particular, the two-triplon contribution should be 
accessible by experiment because it is the leading contribution with an 
even parity.

\subsection{One-Triplon Contribution} 

\begin{figure}[t]
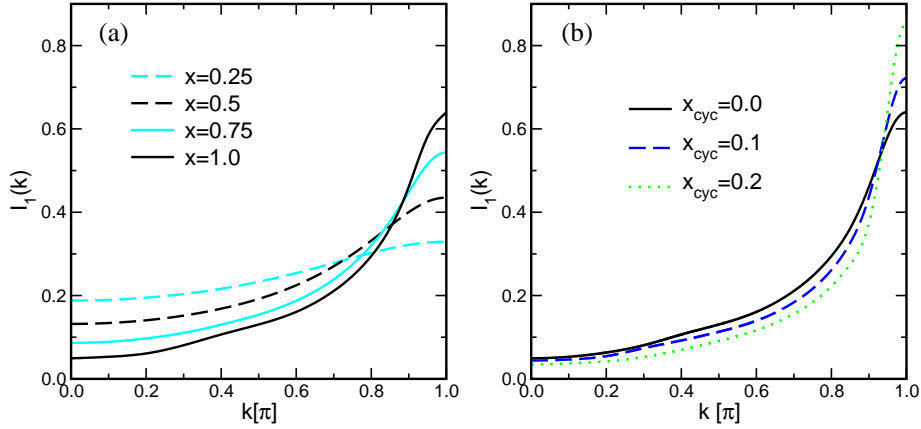

\begin{center}
\includegraphics[width=6cm]{./fig5a.eps}
\includegraphics[width=6cm]{./fig5b.eps}
\end{center}
\caption{(color online) The one-triplon spectral density $I_1(k)$ for 
  ${\mathcal O}^{\rm
    III}$ is shown for (a) $x_{\rm cyc}=0$ and various values of
  $x=\{0.25;0.5;0.75;1.0\}$ and for (b) $x=1$ and various values of
  $x_{\rm cyc}=\{0;0.1;0.2\}$.\label{fig:Asq_4sp}}
\end{figure}

The most important contribution to the dynamic structure factor is the
 one-triplon contribution. The spectral weight  accumulates
around $k=\pi$. Upon increasing $x$, the spectral weight decreases for small
momenta and concentrates more and more at $k=\pi$ as illustrated in
Fig.\ \ref{fig:Asq_4sp}a.

The effect of the four-spin interaction on the one-triplon spectral weight is
exemplified in Fig.\ \ref{fig:Asq_4sp}b for $x=1$ with $x_{\rm cyc}=0$
(solid black line), $x_{\rm cyc}=0.1$ (dashed (blue) line) and 
$x_{\rm cyc}=0.2$ (dotted gray (green) line). 
It can be seen clearly in Fig.\ \ref{fig:Asq_4sp}b that 
the effect of $x_{\rm cyc}$ on the $k$-resolved one-triplon spectral weight is 
similar to the effect of $x$. The spectral weight is reduced at small momenta 
but it increases around $k=\pi$. These findings suggest that the one-triplon 
spectral weight close to  the phase transition to the out-of-phase dimerization
 is governed by the physics at $k=\pi$. 
Similar results were also found by exact diagonalisation\cite{haga02}.

\subsection{Two-Triplon Contribution} 

The two-triplon contribution to the dynamic structure factor is the leading 
part with even parity. Its total spectral weight increases slightly by 
turning on $x_{\rm cyc}$ (see Sect.\ \ref{sec:sw_ladder_4sp}). The generic 
relative spectral weight is about 20\%  to 30\% of the leading 
one-triplon contribution for realistic values of cuprate ladders.
\begin{figure}
  \begin{center}
    \includegraphics[width=7cm]{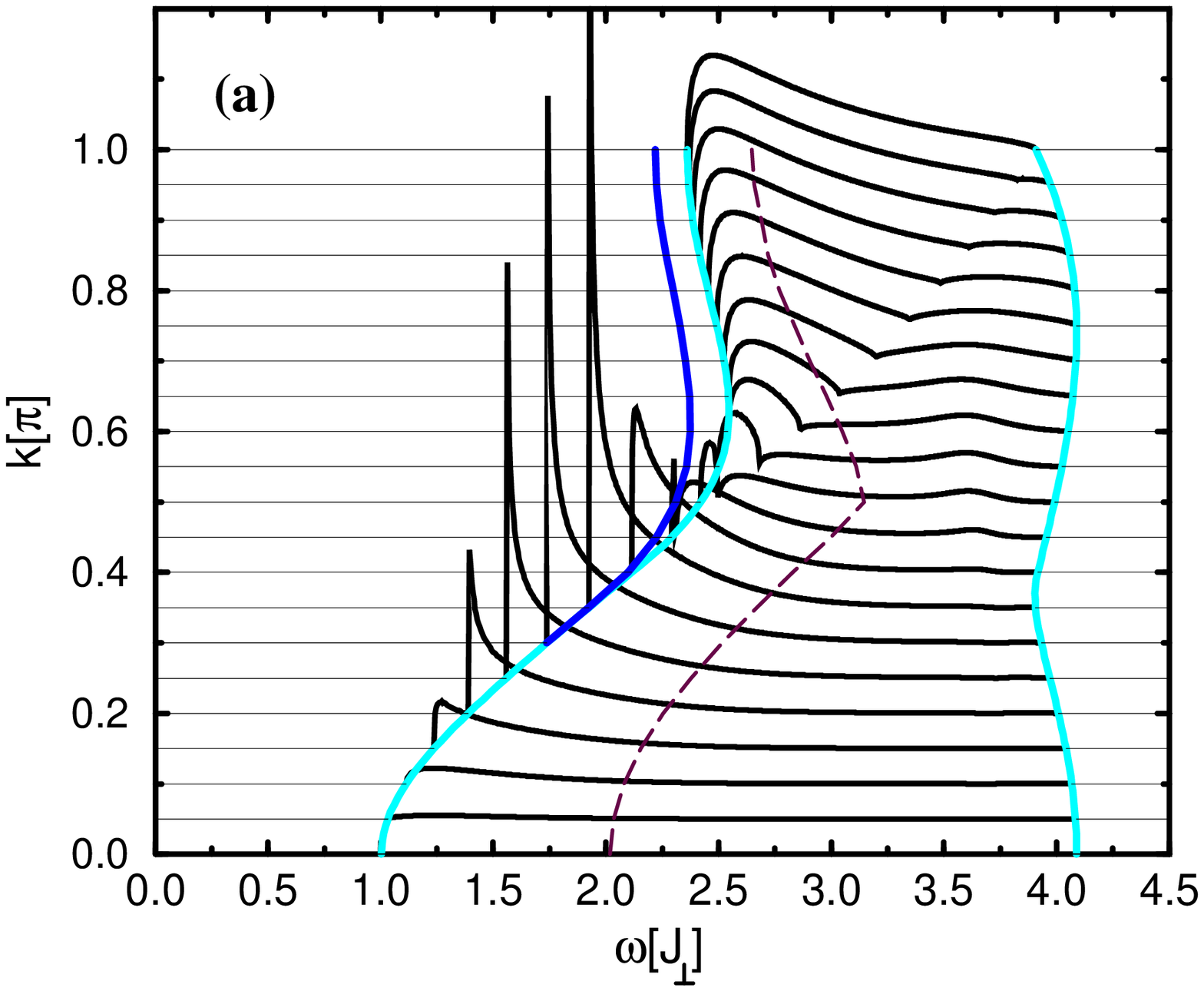}
    \hspace*{-5mm}
    \includegraphics[width=7cm]{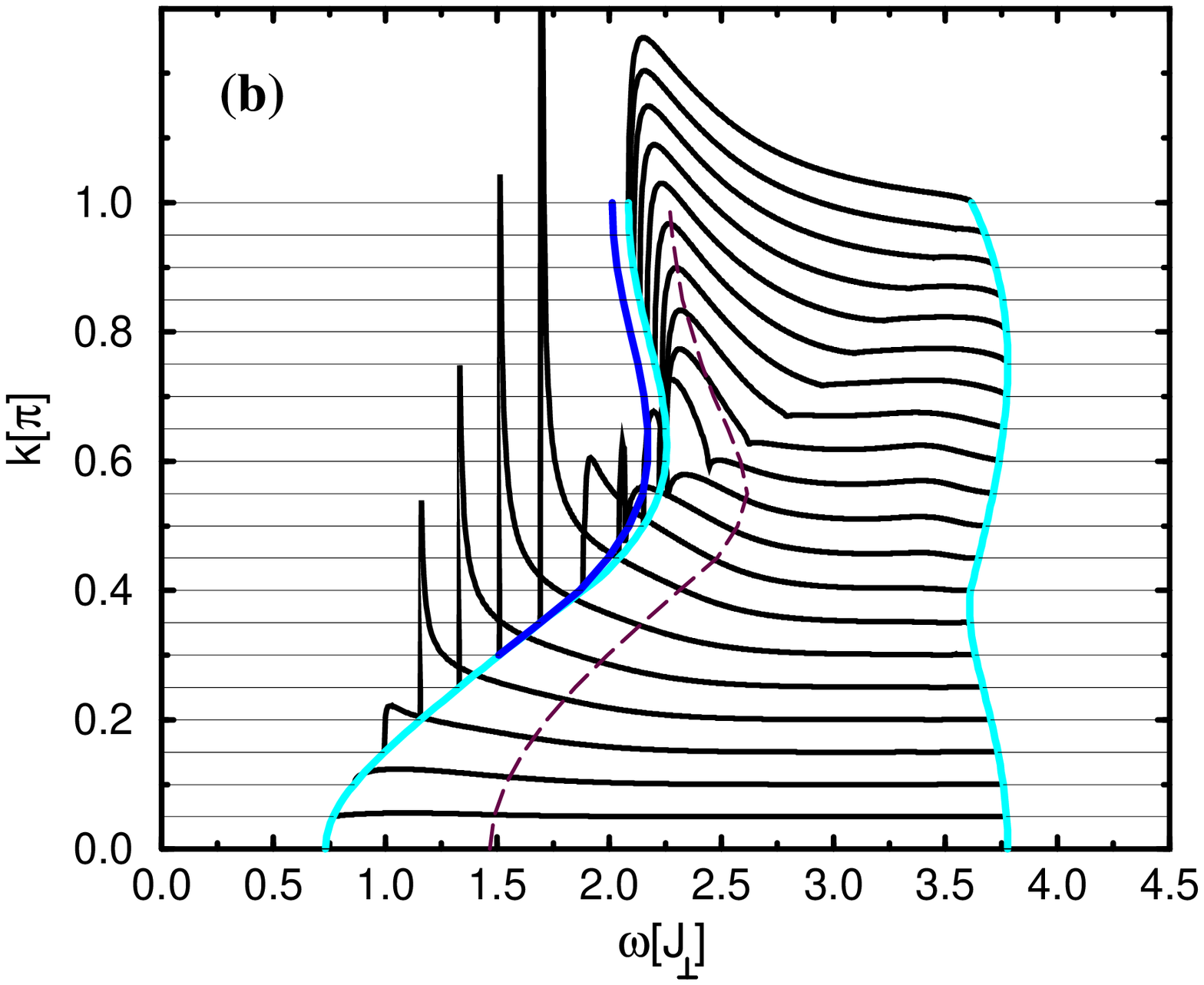}
    \includegraphics[width=7cm]{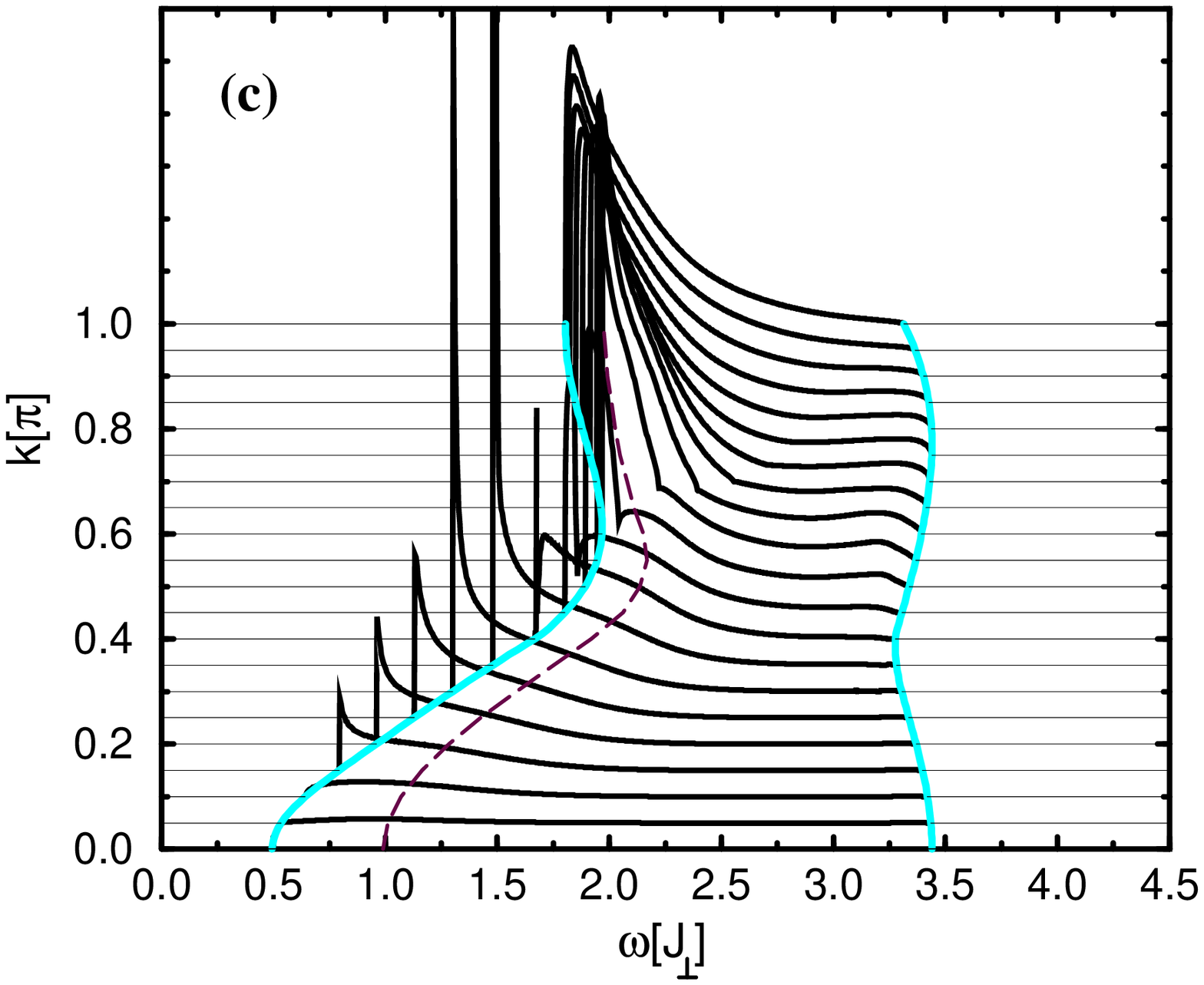} 
    \caption{(color online) 
      The two-triplon spectral density $I_2(k,\omega)$ for 
      ${\mathcal O}^{\rm IV}$ with $x=1.0$ and $x_{\rm cyc}=0.0$ (a), 
      $x_{\rm cyc}=0.1$ (b) and $x_{\rm cyc}=0.2$ (c). The gray (cyan) lines 
      denote the lower and the upper edge of the two-triplon continuum. The 
      black lines indicate the dispersion of the $S=1$ two-triplon bound state.
      The long-dashed dark lines depict the lower edge of the $S=1$ 
      four-triplon continuum.} 
    \label{fig:Spectral_D_S1_4sp}
  \end{center}
\end{figure}

In Fig.\ \ref{fig:Spectral_D_S1_4sp} the result for the two-triplon 
contribution to the dynamic structure factor is shown for $x=1$ and 
$x_{\rm cyc}=0$ (a), $x_{\rm cyc}=0.1$ (b) and $x_{\rm cyc}=0.2$ (c). As 
discussed in Sect.\ \ref{ssec:2tc_4sp} the four-spin ring interaction 
has a strong influence on the $S=1$ two-triplon bound state. The attractive
interaction is lowered by the four-spin interaction which induces
interesting changes in the line shape of the two-triplon spectral density.

\begin{figure*}
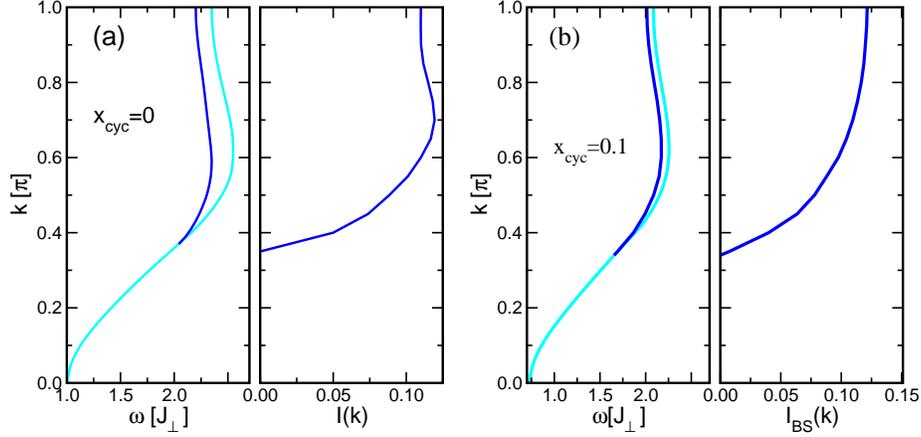

  \begin{center}
    \includegraphics[width=6cm]{./fig7a.eps}
    \includegraphics[width=6cm]{./fig7b.eps}
  \end{center}
  \caption{(color online)
    The $S=1$ two-triplon  bound state energy and the spectral weight 
    for $x=1.0$ with (a) $x_{\rm cyc}=0$ and with (b) $x_{\rm cyc}=0.1$. 
    Left panels: The gray (cyan) line denotes the lower edge of 
    the two-triplon 
    continuum and the black line depicts the dispersion of the 
    $S=1$ two-triplon bound state. Right panels: The momentum-resolved 
    spectral weight of the $S=1$ two-triplon bound state is shown.} 
\label{fig:ladder_BS_S1_4sp}
\end{figure*}

The dispersion of the $S=1$ two-triplon bound state is denoted as a solid 
black line in Fig.\ \ref{fig:Spectral_D_S1_4sp} and in the left panels of 
Fig.\  \ref{fig:ladder_BS_S1_4sp}. In the right panels of the latter figure the
$k$-resolved spectral weight of the bound state is depicted. On increasing
$x_{\rm cyc}$, the binding energy and the spectral weight of the bound state is
reduced. At $x_{\rm cyc}=0.2$, the $S=1$ two-triplon bound state has 
disappeared because the overall attractive interaction has become too small.
This finding makes a clear and directly verifiable
prediction for the results of future INS investigations.
While a spin ladder without ring exchange shows a bound state also
in the $S=1$ channel\cite{knett01b}, 
the spin ladder with ring exchange $x_{\rm cyc}=0.2$ does not.

The reduction of the binding energy of the two-triplon bound state results in
a sharpening of the two-triplon
spectral density at the lower band edge. The effect is
most prominent near $k=\pi$. At the value of  $x_{\rm cyc}$ where the bound 
state is degenerate with the lower edge of the two-triplon continuum, the 
spectral density  displays an inverse square root divergence. 
This corresponds qualitatively to our findings for the
dimerized and frustrated spin-chain at certain values of the exchange 
couplings\cite{schmi04a}. At $x_{\rm cyc}=0.2$, where the two-triplon bound 
state has already dissolved in the two-triplon continuum, 
the very sharp structures near the lower band edge are the remaining
signature of this divergence.

The lower band edge of the $S=1$ four-triplon continuum is depicted by 
long dashed lines in Fig.\ \ref{fig:Spectral_D_S1_4sp}. It can be seen that the
 overlap of the two-triplon and the four-triplon continuum increases with 
increasing four-spin interaction. Possible life-time effects 
are expected to be of minor importance because the spectral weight of the four-triplon continuum increases only very slowly like $I_4 (\omega)\propto (\Delta\omega)^{13/2}$ at the lower band edge\cite{kirsc04} where $\Delta\omega$ measures the distance to the band edge. This extremely slow increase results from the hardcore repulsion of the triplons which makes them behave like fermions at low energies because the system is one-dimensional.

\section{Optical Experiments}
\label{sec_lad_SD_S0}

In this section, two-triplon spectral densities with total spin $S=0$ are 
analyzed. The effect of the four-spin interaction on the energy spectrum was 
already discussed in Sect.\ \ref{ssec:2tc_4sp}. In contrast to the $S=1$ case 
studied in the last section, the $S=0$ two-triplon bound state is present also 
for  $x_{\rm cyc}=0.2$ in the range of parameters considered here. 
Thus, the impact of the four-spin interaction on the line shape of the 
two-triplon spectral density is less pronounced than in the $S=1$ case. 

In the sequel, the complete two-triplon contribution is discussed focusing on  
$x=1$. The general trends are the same for other values of $x$. 
The  second and the third part deal with the two-triplon contribution to 
Raman lines and IR absorption.  In all parts, results are presented for the observable
${\mathcal O}^{\rm I}$ acting on the rungs and  for the
observable ${\mathcal O}^{\rm II}$ acting on the legs of the spin ladder.

\subsection{Two-Triplon Contribution}

The two-triplon contribution comprises most of the spectral weight in the 
$S=0$  sector (see Sect.\ \ref{sec:sw_ladder_4sp}). Nevertheless, the three- 
and  four-triplon spectral weights are sizable for $x\geq 1$ and especially 
for finite $x_{\rm cyc}$. The latter contributions have to be kept in mind when
comparing the results to experimental data and in estimating possible neglected
 life-time effects. Recall that the observable ${\mathcal O}^{\rm I}$ has 
only contributions with an even  number of triplons while the observable
 ${\mathcal O}^{\rm II}$ includes contributions of odd and even number of
triplons.

\begin{figure}
  \begin{center}
    \includegraphics[width=10cm]{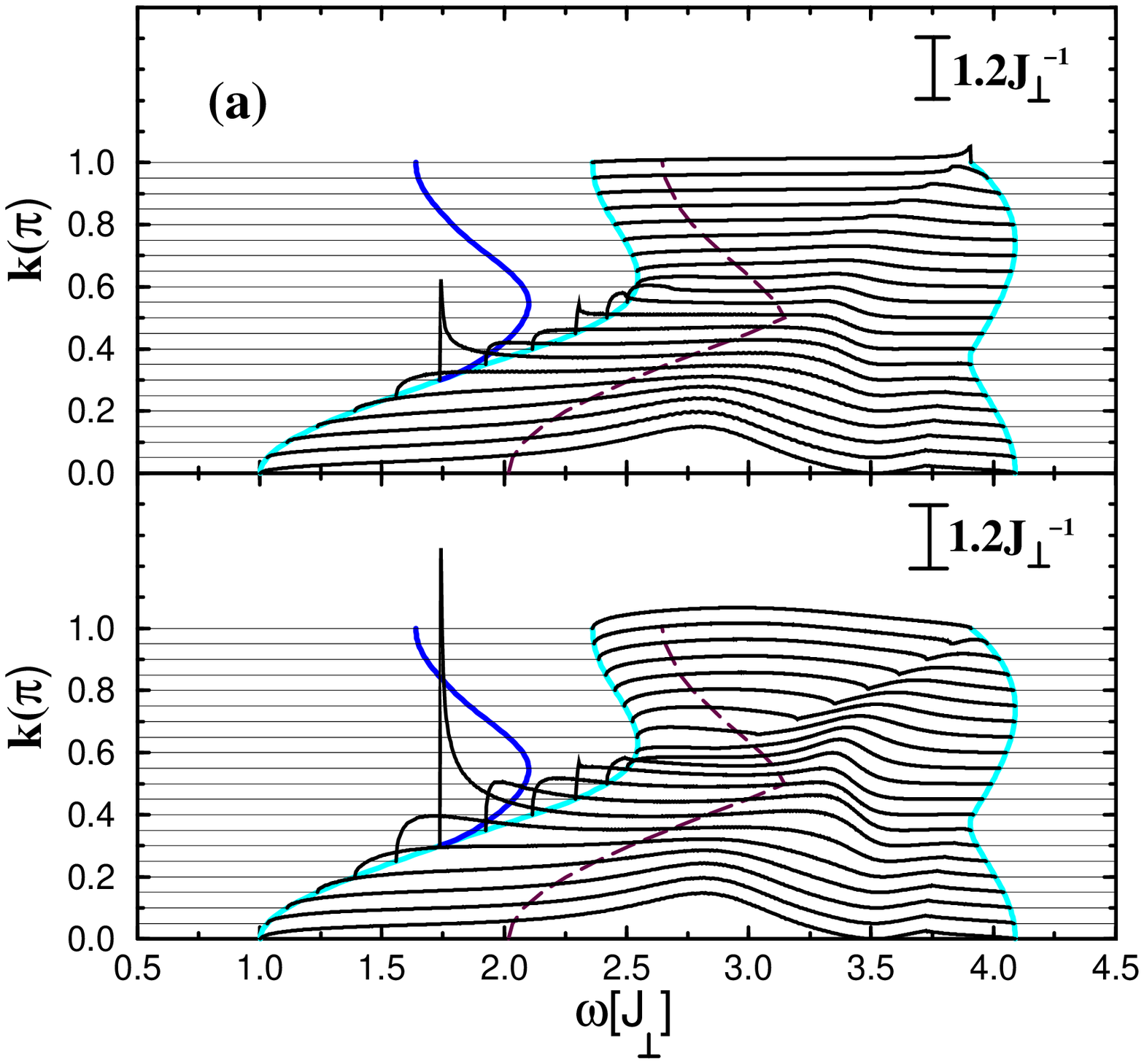}
    \includegraphics[width=10cm]{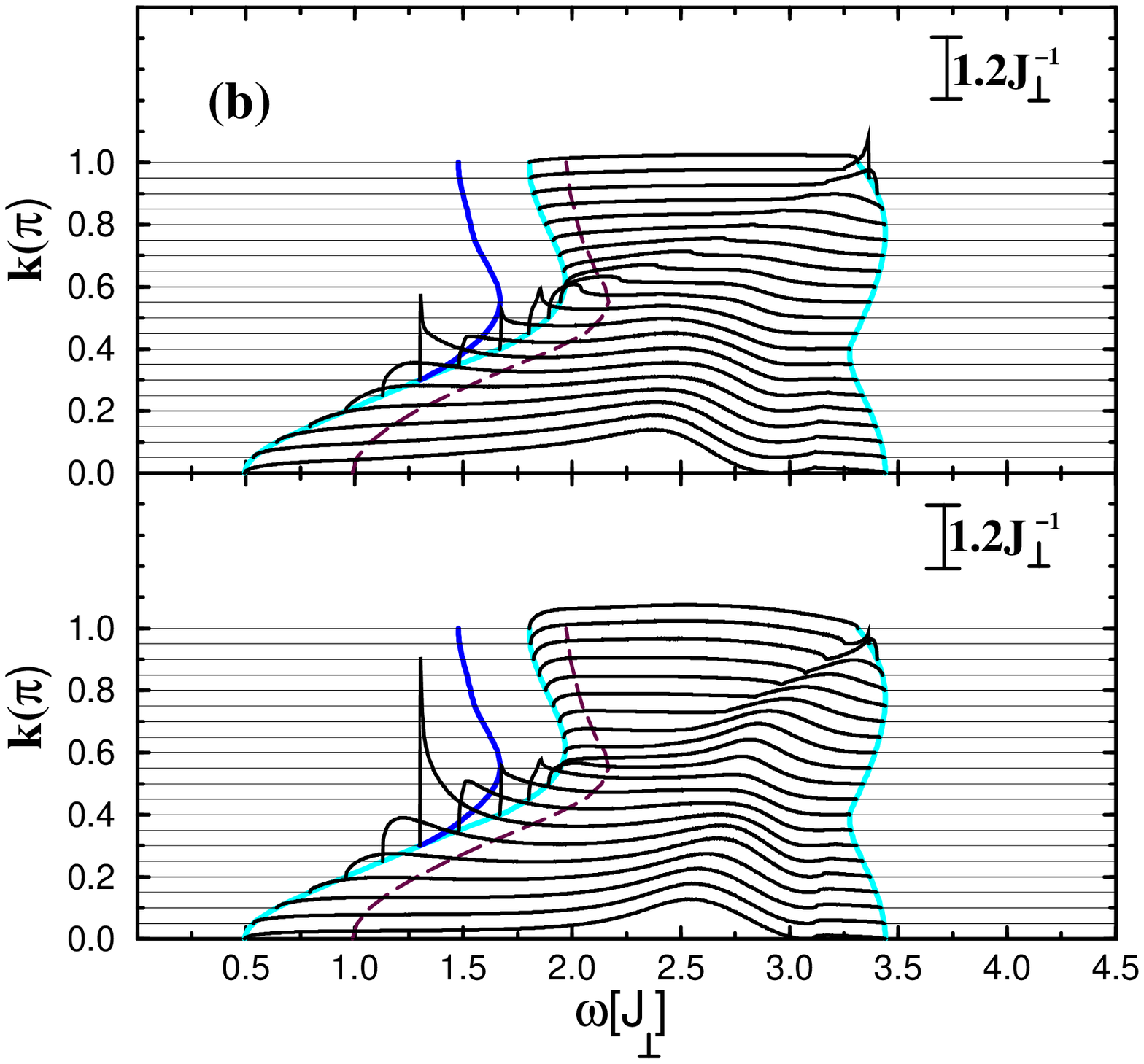}
    \caption{(color online)
      Two-triplon spectral density $I_2(k,\omega)$ with $S=0$ at 
      $x=1.0$ with $x_{\rm cyc}=0.0$ (a) and with $x_{\rm cyc}=0.2$ (b). 
      The upper panels correspond to ${\mathcal O}^{\rm I}$ and the lower 
      panels to ${\mathcal O}^{\rm II}$. The gray (cyan) lines denote the 
      lower and 
      the upper edge of the two-triplon continuum. The black (blue) lines 
      indicate the
      dispersion of the $S=0$ two-triplon bound state. Long-dashed 
      lines depict the lower edge of the $S=0$ four-triplon continuum.} 
    \label{fig:Spectral_D_S0_4sp}
  \end{center}
\end{figure}
\begin{figure*}
  \begin{center}
    \includegraphics[width=8cm]{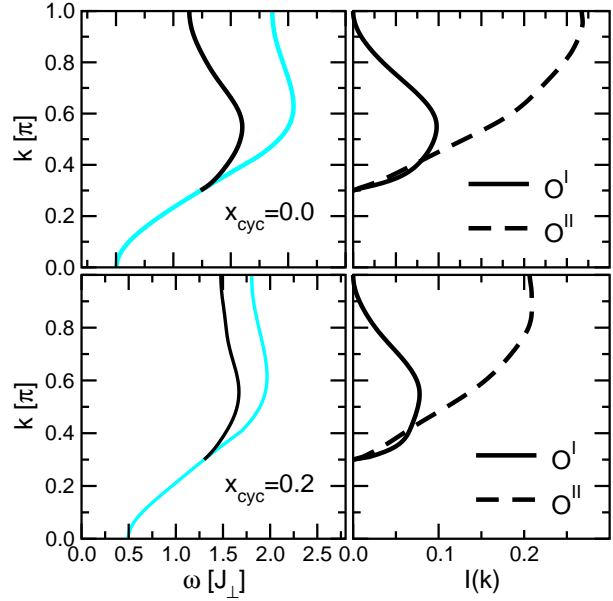} 
  \end{center}
  \caption{(color online)
    Energy and spectral weight of the two-triplon $S=0$ bound states 
    for $x=1.0$ with $x_{\rm cyc}=0.0$ (upper panel) and $x=1.0$ with 
    $x_{\rm cyc}=0.2$ (lower panel). Left panels: The solid gray (cyan) line 
    denotes the lower edge of the two-triplon continuum and the solid black 
    line depicts the dispersion of the $S=0$ two-triplon bound state. 
    Right panels: The momentum-resolved spectral weight of the $S=0$ 
    two-triplon bound state is plotted as measured by ${\mathcal O}^{\rm I}$ 
    (solid line) and by ${\mathcal O}^{\rm II}$ (dashed line).} 
  \label{fig:ladder_BS_S0_4sp}
\end{figure*}

In Fig.\ \ref{fig:Spectral_D_S0_4sp} results for $x=1$ with $x_{\rm cyc}=0.0$
(a) and $x_{\rm cyc}=0.2$ (b) are shown. The upper panels depict the findings 
for ${\mathcal O}^{\rm I}$ and
the lower panels depict the results obtained for ${\mathcal O}^{\rm II}$. 
Detailed information about the $S=0$ two-triplon bound state dispersion and its
 $k$-resolved spectral weight is presented in Fig.\ \ref{fig:ladder_BS_S0_4sp}.
The first effect to note is that the overall line shape of the two-triplon 
spectral density hardly changes in the $S=0$ case when the four-spin 
interaction is switched on. As in the $S=1$ sector, there is a global shift to 
lower energies at finite $x_{\rm cyc}$ resulting from the change of the 
one-triplon dispersion.

Both observables yield the same line shape for $k=0$ and $x_{\rm
  cyc}=0$ \cite{freit00b,schmi01}. 
This symmetry holds only at $k=0$; it is broken for 
finite $x_{\rm cyc}$. Nevertheless, even at $x_{\rm cyc}=0.2$ the line shape 
for both observables is very similar for small momenta. 
A more detailed discussion of the $k=0$ 
contribution will be given in the next subsection about Raman scattering.
For large momenta, however, the responses for the two observables differ.

The next important effect of the four-spin interaction besides the global shift
to lower energies is the decrease of the binding energy near 
$k=\pi$. The loss of binding energy can be seen clearly in Fig.\ 
\ref{fig:ladder_BS_S0_4sp}. This effect is accompanied by a reduction of 
spectral weight of the bound state near $k=\pi$, see right panels of Fig.\ 
\ref{fig:ladder_BS_S0_4sp}. The spectral weight is transferred to the 
two-triplon  continuum which becomes sharper on increasing $x_{\rm cyc}$.

The black long-dashed curve in Fig.\ \ref{fig:Spectral_D_S0_4sp} represents the
 lower band edge of the $S=0$ four-triplon continuum. The overlap is slightly enhanced by the four-spin interactions. Note that the 
two-triplon bound state is located below the four-triplon continuum for all 
momenta so that no finite life-time effects appear for this 
part of the two-triplon contribution. Furthermore, the two-triplon contribution
 displays  only broad features for large momenta and energies above the long 
dashed curve. Therefore life-time effects will have only minor effects on this 
part of the spectrum because a small additional broadening will not change 
much.
  
\subsection{Raman Spectroscopy}

This subsection deals with the two-triplon contribution to the non-resonant 
Raman response. The observables for magnetic light scattering in rung-rung 
(leg-leg) polarization are the $k=0$ part of ${\mathcal O}^{\rm I}$ 
(${\mathcal O}^{\rm  II}$) 
\cite{schmi01,schmi03b,freit00b,schmi05a,fleur68,shast90b}. 
Here only a short summary of the most important results is given. The reader
interested in further details is referred to more specialized 
papers\cite{schmi01,schmi03b,schmi05a}.

Channels with an odd number of triplons are inaccessible by
Raman scattering due to the invariance of the observables
with respect to reflection about the centerline of the ladder,
see discussion on the parity in Sect.\ \ref{sect:method}. Thus only
excitations with an even number of triplons matter. The relevant 
contributions to the Raman response come from the two-triplon and the 
four-triplon sector. The two-triplon contribution is the most important one,
cf.\ Sect.\ \ref{sec:sw_ladder_4sp}.

The line shape does not depend on the observable for $x_{\rm cyc}=0$ 
because the Hamiltonian is a weighted sum of the two of them $H =
{\mathcal O}^{\rm I}(k=0)+x{\mathcal O}^{\rm II}(k=0)$ up to a global
factor\cite{schmi01,freit00b}. Thus the excited state ${\mathcal
 O}^{\rm I}(k=0)|0\rangle$ equals $-x{\mathcal O}^{\rm II}(k=0)|0\rangle$ 
except for a component proportional to the ground state $|0\rangle$ which does 
not represent an excitation. The Raman response for $x_{\rm cyc}=0$ is 
discussed in detail elsewhere\cite{schmi01,schmi05a}.
The line shape of $\mathcal{O}^{\rm I}(k=0)$ and ${\mathcal O}^{\rm II}(k=0)$
for finite $x_{\rm cyc}$ is no longer the same.
We find that the line shape of $\mathcal{O}^{\rm I}(k=0)$ is 
almost unchanged by switching on $x_{\rm cyc}$. In contrast, the line shape of 
${\mathcal O}^{\rm II}(k=0)$ changes although
the effect is not large. The overall difference 
between the line shapes in both polarizations remains small for the 
realistic parameters discussed in this work. The next important effect
 of the four-spin interaction besides  the global shift to lower frequencies is
 a sharpening of the dominant two-triplon peak measured by 
${\mathcal O}^{\rm II}(k=0)$, i.e., in leg-leg polarization. In addition, an 
almost constant plateau is produced for freqencies smaller than the two-triplon
 peak in this polarization. 

\subsection{IR absorption}

In this section a short discussion of phonon-assisted IR absorption of magnetic
excitations is presented. The leading infrared-active magnetic absorption is a 
two-triplon-plus-phonon process\cite{windt01,loren95a,loren95b}. Note that 
three- and four-triplon processes contribute also to the IR absorption signal,
depending on the exchange couplings of the ladder, 
see Fig.\ \ref{fig:sw_4sp}b. As in the previous
subsection, we concentrate here on the most important and generic features 
and we refer to other publications for more detailed 
discussions\cite{nunne02,windt01,gruni02b,gruni03a}.

The two-triplon spectral density $I_2 (k,\omega)$ has to be integrated over
all momenta weighted by a phonon-specific form factor $|f_{\rm ph}(k)|^2$.
This yields the two-triplon part of the IR absorption $I^{\rm IR}_{\rm 2trp}$. The 
precise form of the phonon form factor $|f_{\rm ph}(k)|^2$ depends on the
specific phonon involved\cite{nunne03}.
Different phonons have different form factors so that the sum
of several contributions has to be considered. 
But usually all relevant phonon form factors are largest for large momenta
at the zone boundary.

 In leg polarization, three features are present in the experimental
signals: two peaks at lower energy and a broad hump at higher energies. 
At lower energy, the  line shape is dominated by the two-triplon $S=0$ bound 
state as predicted for small values of $x$.\cite{jurec00} 
For $x\gtrapprox 0.5$, the 
 dispersion of the bound state displays a maximum at $k\approx \pi/2$ and a 
 minimum at $k=\pi$, see Sect.\ \ref{ssec:2tc_4sp}. 
 Both give rise to van Hove singularities in the density
 of states which cause the peaks in $I^{\rm IR}_{\rm 2trp}$ \cite{windt01}. 
 The spectral weight of the bound state has a maximum at $k=\pi$ for 
 ${\mathcal O}^{\rm II}$, see Fig.\ \ref{fig:ladder_BS_S0_4sp}. 
 Thus both extrema, the minimum and the maximum, lead to 
peaks in the two-triplon contribution in leg polarization which
originate from the broadened inverse square root 
van Hove singularities in one dimension.
 The intensity of the energetically  lower peak is larger. The peaks are more 
 clearly seen for larger values of  $x$ because the difference between maximum 
 and minimum of the two-triplon bound state dispersion increases with 
 increasing $x$. The third feature, the broad hump, originates from the 
 two-triplon continuum and from three-triplon 
contributions\cite{nunne02,kirsc04}.

In rung polarization, the three features are also present but the intensity is 
distributed differently. The spectral weight of the two-triplon bound 
state vanishes at $k=\pi$ (Fig.\ \ref{fig:ladder_BS_S0_4sp}). Thus the 
energetically lowest van Hove singularity appears only as low-energy shoulder.
It becomes more  pronounced for larger $x$ values. The dominant peak results 
from the van Hove singularity at $k\approx\pi/2$ where the spectral weight
is finite. The broad hump induced by the two-triplon continuum 
is relatively more important in this polarization because the two singularities at lower energies carry less weight.

In leg polarization, channels with both odd and even number of triplons
contribute. The analysis of the spectral weights, see Sect.\ 
\ref{sec:sw_ladder_4sp}, has shown that 
the three-triplon channel contributes significantly. This was to be
expected from earlier DMRG calculations\cite{nunne02}.
Due to the reflection symmetry of the observable, contributions with odd 
parity are absent in rung polarization. Therefore, 
the next important correction is only the four-triplon contribution 
which carries little weight, see Sect.\ \ref{sec:sw_ladder_4sp}.
Moreover, its line shape can be expected to be rather featureless and broad,
because as many as four particles are involved so that momentum 
conservation does not have a very important impact.
 
\section{Summary and Conclusion}
\label{sect:sum}

In this article the spectral properties of magnetic excitations in cuprate
two-leg ladder systems have been reviewed and extended to finite ring exchange.
 A quantitative microscopic description comprises
two-spin and four-spin interactions. The undoped cuprate ladders are always in
the rung-singlet phase, i.e. there is an adiabatic connection to the state of
isolated rung dimers. The elementary excitation is a triplon. The exchange
couplings generically take the values $x=J_\parallel
/J_\perp =1.2-1.5$, $x_{\rm cyc}/J_{\perp} \approx 0.2$ and $J_\perp =
1000-1200$cm$^{-1}$.

The accuracy of the description in terms of rung triplons in the regime 
relevant for the cuprate ladders was discussed by means of the spectral 
weights. It has been found that the leading one- and two-triplon channels 
contain the dominant weight. A sizable three-triplon contribution is found for 
total spin $S=0$. The one- and two-triplon channels dominate the physics for
realistic values of the coupling constants for cuprate ladders.

Various energies are important: the one-triplon dispersion, the
energies of bound states consisting of two triplons and the positions of
the multi-triplon continua. The main effect of
the four-spin interaction is a global red shift to lower energies. As a
consequence, the continua approach each other. The two- and the
four-triplon continua especially overlap more strongly. 
The second sizable effect of the 
four-spin interaction is to lower the attractive two-triplon interaction. 
Thereby the binding energy of the two-triplon bound states is reduced and 
spectral weight is shifted from the bound state to
 the two-triplon continuum. The $S=1$ bound state has disappeared for 
realistic  values of cuprate ladders due to this effect.

The dynamic structure factor is dominated by the one-triplon
contribution. Most of the spectral weight is found at the zone boundaries
where the spin gap occurs. 
The spectral weight for small momenta is very low. The two-triplon
contribution represents about 30\% of the total spectral weight
integrated over all momenta, frequencies and parities. The two-triplon
contribution is the leading one with even parity. Thus it should be
measurable independently from the one-triplon contributions. 
The shape of the two-triplon  continuum for total spin $S=1$
changes significantly for finite four-spin interaction. The disappearance of
the two-triplon bound state causes sharp structures at the lower band edge of
the two-triplon continuum.

The spectral densities with total spin $S=0$  are relevant for Raman scattering
and IR absorption. The electromagnetic waves modify the exchange couplings on the rungs 
\emph{or} on the legs of the ladder, respectively,
 leading to a pronounced dependence on the polarization. Raman scattering
 measures the response at zero total momentum. There the spectrum is dominated 
by the two-triplon peak within the two-triplon continuum. The IR absorption displays 
a structure with three peaks. The two low-energy features originate
from the two-triplon bound state. The third very broad peak represents the
continuum contribution. The overall shape of the $S=0$ spectral density is 
not changed significantly for realistic values of the four-spin interaction.

The interplay of recent developments in the theoretical description of gapped 
one-dimensional spin liquids and the improvements in spectroscopic experiments 
have permitted to gain a quantitative understanding of the
magnetic excitations and their line shapes in cuprate ladder systems.

Fruitful discussions are acknowledged with M. Gr\"uninger, A. G\"o\ss{}ling,
A. L\"auchli, A. Reischl, S. Dusuel, S. Trebst, H. Monien, D. Khomskii, and E. 
M\"uller-Hartmann. 
This work was supported by the DFG in SFB 608 and in SP 1073.


\end{document}